\documentclass[a4paper,10pt]{article}
\usepackage[utf8]{inputenc}
\usepackage{amsmath,amsfonts,amssymb}

\title{Time Crystals from Minimum Time Uncertainty }
 
\author{Mir Faizal$^a$\footnote{f2mir@uwaterloo.ca} , Mohammed M. Khalil$^b$\footnote{moh.m.khalil@gmail.com} , 
Saurya Das$^c$\footnote{saurya.das@uleth.ca} \\
\\
$^a$Department of Physics and Astronomy,\\ University of Waterloo, Waterloo,
Ontario, N2L 3G1, Canada\\
{} \\
$^b$ Department of Electrical Engineering,\\
Alexandria University, Alexandria 12544, Egypt\\
{}\\
 $^c$ Department of Physics and Astronomy,\\
University of Lethbridge, 4401 University Drive,\\
Lethbridge, Alberta, T1K 3M4, Canada  
}
\date{}
\begin{document}

\maketitle 
\begin{abstract}
Motivated by the Generalized Uncertainty Principle, covariance,
and a minimum measurable time, we propose a deformation of the Heisenberg algebra and show that 
this leads to corrections to all quantum mechanical systems. 
We also demonstrate that such a deformation implies a discrete spectrum for time. In other words, time 
behaves like a crystal. As an application of our formalism, we analyze the effect of such a deformation on the rate of spontaneous emission in a hydrogen atom. 
\end{abstract}

\section{Introduction}

The Heisenberg uncertainty principle predicts that the position of a particle can, in principle, be measured as accurately as
one wants if its momentum is allowed to remain completely uncertain. 
However, most approaches to quantum gravity 
predict the existence of a minimum  measurable length scale, 
usually the Planck length. 
There are also strong indications from
black hole physics and other sources 
for the existence of a minimum measurable length 
\cite{z4,z5,hoss}. This is because  the energy needed to probe spacetime below the Planck length scale exceeds 
the energy needed to produce a 
black hole in that region of spacetime.
Similarly, string theory also predicts a minimum length,
as strings are the smallest probes \cite{z2,zasaqsw,csdcas,cscds,2z}. 
Also in loop quantum gravity there exists a  minimum measurable 
length scale, which  
turns the big bang into a big bounce  \cite{z1}.

The existence of a minimum measurable length scale in turn
requires the modification of the Heisenberg uncertainty principle into a Generalized Uncertainty Principle
(GUP) \cite{z2,zasaqsw,csdcas,cscds}; there is a 
corresponding deformation of the Heisenberg algebra
to include momentum-dependent terms, and a modified 
coordinate representation of the momentum operators \cite{2z,14,17,18,5,51,54}.
It may be noted that a different kind of deformation of the 
Heisenberg algebra occurs due to Doubly Special Relativity (DSR) theories, which postulate the existence of a 
universal energy scale (the Planck scale) 
\cite{2,21,3}. 
These are also related to the idea of 
discrete spacetime \cite{1q}, spontaneous symmetry breaking of Lorentz invariance in string
field theory \cite{2q}, spacetime foam models \cite{3q}, spin-network in loop quantum gravity \cite{4q}, non-commutative geometry \cite{5q, q5, 55}, ghost condensation in perturbative quantum gravity \cite{7q}, and 
Horava--Lifshitz gravity \cite{6q}. It may be noted that DSR has been generalized to curved spacetime and the resultant theory is called 
gravity's rainbow \cite{n1, n2, n0aa, n4aa, n5aa, n6aa}. It is
interesting  to note that the  deformation from DSR and the deformation from GUP can be combined 
into a single consistent 
deformation of the Heisenberg algebra \cite{main2}. 

A number of interesting quantum systems 
have been studied using this deformed algebra, such as 
the transition rate of ultra-cold neutrons in gravitational field  
\cite{n6}, the Lamb shift and Landau levels \cite{n7}. 
There has been another interesting result derived from this deformed algebra, which shows that space needs to
be a discrete lattice, and only multiples of a fundamental
length scale (normally taken as the Planck length) can be measured \cite{m1}. Note that minimum length
does not automatically imply discrete lengths,
or vice versa. 
Motivated by this result, in this paper we 
analyze the deformation of the algebra and the 
subsequent 
Schr\"{o}dinger equation consistent with the existence of a minimum time, and demonstrate that it leads 
to a discretization of time as well.
It may be noted that discretization of time had also been
predicted from a deformed version of the Wheeler--DeWitt equation \cite{wd}. 
%
The discretization of time, 
and the related 
breakdown of time reparametrization invariance of a system
resembles a crystal lattice in time.
%
%
%
Time crystals have been studied recently using a very different physical motivation, 
e.g. analyzing superconducting rings, and the
spontaneous breakdown of time-translation symmetry in 
classical and quantum systems 
\cite{ctime,  ctime1, c0time, c1time, cti}.

\section{Observable Time}
 In this section, we will review the  work done on viewing   time as a quantum mechanical observable. 
It is well known that time cannot be represented as a self-adjoint operator 
\cite{1a}. This is because the Hamiltonian with a semi-bounded spectrum does not admit a group 
of shifts which can be generated from  canonically conjugate self-adjoint operators. 
However,  von Neumann had suggested that restricting quantum mechanics to 
self-adjoint operators could be quite limiting \cite{vo}. In fact, it was demonstrated 
by von Neumann that  the momentum operator for a
free particle bounded by a rigid wall 
at $x = 0$ is  not a self-adjoint operator but only a maximal Hermitian operator. 
This  situation is similar to the time being defined as an observable.

It has been demonstrated 
that under certain conditions time can be viewed as a quantum mechanical observable
\cite{time, sdfs, dscsdc, svdcvs, time1}. 
This is because  it is possible to use symmetric non-self-adjoint operators that  satisfy 
the   commutation relation  \cite{2a,a2}, 
\begin{equation}
[t, H] = - i \hbar  
\end{equation}
In this formalism, observables are
viewed as positive operator valued  measures. 
Now for  a system with Hamiltonian 
$H$ the  map $b \to e^{iHb}$ constitutes a unitary representation of the time translation group.  
Thus, the positive operator valued $B$, with $\theta \to B(\theta) $ represents a time observation of the system, 
and it will satisfy $ e^{iHb} B(\theta) e^{-iHb} = B(\theta - b)$. So for a time observable $B$, it is possible to define 
  a symmetric time operator $t = \int t dB(t)$. 
This operator will    not be self-adjoint. However, self-adjointness is not essential for  calculating
probabilities associated with the system. So, for any experiment the probability measure $\theta \to p(\theta)$ 
can be 
be associated with the states $\rho$  by defining $p(\theta) = tr [\rho B(\theta)]$, where 
$\theta \to B(\theta)$ is a positive operator valued  measure \cite{time}. Thus, it is 
possible to formally define time as an 
observable by using a   maximal Hermitian (but non-self-adjoint) operator for time.

It is this definition of time that
we will use when formally  deforming the commutation relation.  What we intend to do in this paper is to deform this 
formal definition of time to be consistent with the existence of a minimum measurable time interval. 
Mathematically this situation will be similar to 
the GUP deformation of the usual Heisenberg algebra.  
Physically observable time can be defined by   defining an 
 observable with   reference to the evolution of some 
non-stationary quantity, if events are characterized 
by   of a  specific values of  this quantity
\cite{time}. Such a   
non-stationary quantity could be the  tunneling time for particles. 
Then the existence of a minimum measurable time interval 
will constitute a lower bound on such measurements. 
The existence of a lower bound on such measurements will effect the measurement of 
 tunneling time for particles. In fact, such system have been analyzed by considering 
time as an observable \cite{sdfs, dscsdc, svdcvs, time1}. Even though such an analysis is important, 
we will concentrate on another problem in this paper. We will analyze   the 
deformation of commutator between the Hamiltonian and time, and demonstrate that such a deformation can 
lead to the existence of a discrete spectrum for time.

\section{Minimum Time}

We start with the modified Heisenberg algebra, the 
modified expression of the momentum operator in 
position space, 
and the GUP consistent with all theoretical models, correct to ${\cal O}(\alpha^2)$. In this paper, we use units in which $c=1$. We have
\begin{eqnarray}
[x^i, p_j] &=&  i \hbar \left[  \delta_{j}^i - \alpha |p^k p_k|^{1/2} \delta_{j}^i + \alpha |p^k p_k|^{-1/2} p^i p_j
 \right.  \nonumber \\  && \left. + \alpha^2 p^k p_k \delta_{j}^i + 3 \alpha^2 p^i p_j\right], \\
p_i &=& -i \hbar\left(1 -  \hbar\alpha \sqrt{- \partial  ^j \partial_j} - 2\hbar^2 \alpha^2    \partial ^j
\partial _j\right) \partial _i
\label{3}, 
\end{eqnarray}
where $\alpha = {\alpha_0 \ell_{Pl}}/{\hbar}$,
and $\ell_{Pl}$ is the Planck length. 
It  has been suggested that the 
 parameter $\alpha_0$ could
be situated at an intermediate scale  between the electroweak scale and the
Planck scale, and this could have measurable consequences in the near future  \cite{n7}. 
However, if  such a deformation parameter exists, then it would be universal for all processes. This is because it would be the parameter controlling low energy phenomena  occurring because of quantum gravitational   effects, and as gravity affects all systems universally, we expect this parameter also to   universally deform all quantum mechanical systems. 
Also the apparent non-local nature of operators in Eq. \eqref{3} above poses no problem 
in one dimension (space or time). 
In more than one dimensions, the issue was tackled by using
the Dirac equation \cite{main2}. It is also possible to deal with these non-local derivatives,
in more than one dimensions,  using the theory of harmonic extension of functions \cite{mf00, mf22}.
The modified Heisenberg algebra is consistent with the following 
GUP, in one dimension \cite{n7}:
\begin{eqnarray}
\Delta x \Delta p &\geq& 
\frac{\hbar}{2} \left[
1 - 2\alpha \langle p \rangle + 4\alpha^2 \langle p^2 \rangle 
\right] \nonumber \\
&\geq& \frac{\hbar}{2} \left[
1 + \left(
\frac{\alpha}{\sqrt{\langle p^2\rangle}} 
+ 4 \alpha^2  
\right) \Delta p^2 + 4\alpha^2 \langle p\rangle^2
-2\alpha \sqrt{\langle p^2 \rangle}
\right].
\end{eqnarray}
One way to arrive at the temporal deformation of the commutator is to use 
the principle of covariance and propose the following  deformation 
spacetime commutators:
\begin{eqnarray}
 [x^\mu, p_\nu] &=& i \hbar \left[  \delta_{\nu}^\mu - \alpha |p^\rho p_\rho|^{1/2} \delta_{\nu}^\nu + \alpha |p^\rho p_\rho|^{-1/2} p^\mu p_\nu
\right.  \nonumber \\  && \left. + \alpha^2 p^\rho p_\rho \delta_{\nu}^\mu + 3 \alpha^2 p^\mu p_\nu\right], \\
p_\mu &=& -i \hbar \left(1 -  \hbar \alpha \sqrt{- \partial  ^\nu \partial_\nu} - 2 \hbar^2 \alpha^2  \partial ^\nu
\partial _\nu\right) \partial _\mu. \label{covp1}
\end{eqnarray} 
Even though  we could study a temporally deformed  system by using  the temporal part of this covariant algebra, we will only deform the commutation relation between energy and time. 
This is because  the deformation of the spatial part of the Heisenberg algebra has 
  been thoroughly analyzed \cite{main2, n6, n7, m1}, 
 and here we would like to analyze the effect of temporal deformation alone on  a system. 
We will also simplify our analysis by only   deforming the relation between 
 time and Hamiltonian of a system. This 
 deformation will be different from the temporal part of the 
deformed  covariant algebra. 
 It may be noted that such a deformation only makes sense 
 if we view time as a quantum mechanical observable.
Therefore we first define the original commutator of this observable time with 
Hamiltonian as $[t, H] = - i \hbar  $ \cite{2a, a2}. Then we deform 
  this commutator of the observable time with Hamiltonian to 
\begin{eqnarray}
 [t, H] &=& - i \hbar \left[1  + f(H)\right],
\end{eqnarray}
where $f(H)$ is a suitable function of the Hamiltonian of the system.  
Thus, the temporal part of Eq. \eqref{covp1} yields the 
modified Schr\"{o}dinger equation
\begin{equation}
\label{scht}
H \psi = i\hbar\partial_t\psi+\hbar^2\alpha\partial_t^2\psi. 
\end{equation}
As can be seen from the above, this deformation of quantum Hamiltonian will produce 
corrections to all quantum mechanical 
systems.  
The temporal part also implies the following time-energy uncertainty:
\begin{eqnarray}
\label{teu}
\Delta t \Delta E &\geq& 
\frac{\hbar}{2} \left[
1 - 2\alpha \langle E \rangle + 4\alpha^2 \langle E^2 \rangle 
\right] \nonumber \\
&\geq& \frac{\hbar}{2} \left[
1 + \left(
\frac{\alpha}{\sqrt{\langle E^2\rangle}} 
+ 4 \alpha^2  
\right) \Delta E^2 + 4\alpha^2 \langle E\rangle^2
-2\alpha \sqrt{\langle E^2 \rangle}
\right].
\end{eqnarray}

\section{ Time Crystals}
The spatially deformed  Heisenberg algebra has been used for analyzing a free particle in a box \cite{m1}. 
The boundary conditions 
which were used for analyzing this  system were $\psi(0) =0$ and $\psi(L) =0$, where $L$ was
the length of the box. It was demonstrated that the length of the box was quantized because of the 
 spatial deformation of the   Heisenberg algebra. As this 
particle was used as a test particle to measure the length of the box, this implied that space itself
was quantized. 
The same argument can be  now used for the temporal deformation. This can be done by taking 
the  temporal analog of the particle in a box. The  boundary conditions for this system can be written  
as $\psi(0) =0$ and $\psi(T) =0$, where $T$  is a fixed  interval of time. This is the temporal analog 
of a particle in a box, and the 
particle in this case is a test particle which  measures the 
interval of time. Now we will demonstrate that in this 
case the interval of time has to be quantized. As this particle is a test particle used to measure this interval 
of time, we can argue that time itself is quantized. 
 
The temporal part of the deformed  Schr\"{o}dinger equation to first order in $\alpha$ is given by 
\begin{equation}
\label{schrod}
i\hbar\partial_t\psi+\hbar^2\alpha\partial_t^2\psi = E\psi,
\end{equation}
and it  has the solution
\begin{equation}
\label{psit}
\psi(t)=Ae^{\frac{-it\left(1+\sqrt{1-4E\alpha}\right)}{2\alpha\hbar}}+ Be^{\frac{-it\left(1-\sqrt{1-4E\alpha}\right)}{2\alpha\hbar}}.
\end{equation}
Applying the boundary condition $\psi(0)=0$ leads to $B=-A$, and the second boundary condition $\psi(T)=0$ leads to
\begin{equation}
Ae^{\frac{-iT\left(1+\sqrt{1-4E\alpha}\right)}{2\alpha\hbar}} \left(1-e^{\frac{iT\sqrt{1-4E\alpha}}{\alpha\hbar}}\right)=0,
\end{equation}
which means that either $A=B=0$ or both the real and the imaginary parts of the above equation are zero. The real part is
\begin{equation}
-2\sin\left(\frac{T}{2\alpha\hbar}\right)\sin\left(\frac{T\sqrt{1-4E\alpha}}{2\alpha\hbar}\right)=0.
\end{equation}
The imaginary part is
\begin{equation}
-2\cos\left(\frac{T}{2\alpha\hbar}\right) \sin\left(\frac{T\sqrt{1-4E\alpha}}{2\alpha\hbar}\right)=0.
\end{equation}
If both are zero, then
\begin{equation}
\sin\left(\frac{T\sqrt{1-4E\alpha}}{2\alpha\hbar}\right)=0,
\end{equation}
leading to
\begin{equation}
\frac{T\sqrt{1-4E\alpha}}{2\alpha\hbar}=n\pi,
\end{equation}
where $n\in {Z}$. This means that 
\begin{equation}
\label{tint}
T=n\pi \frac{2\alpha\hbar}{\sqrt{1-4E\alpha}}, 
\end{equation}
or expanding in terms of $\alpha$
\begin{equation}
\label{texp}
T=2n\pi\hbar \left(\alpha+2E\alpha^2+6E^2\alpha^3+{\cal O}(\alpha^4)\right)
\end{equation}
i.e. we can only measure time in discrete steps.
It is interesting to note that this discrete interval is dependent on the energy of the system, 
i.e. the
larger the energy the larger will be this discrete interval of time, but since the energy dependence 
is to second and higher orders, this does not change the time interval by much, except 
near Planckian energy scales. It may also be noted that this time interval is of the same order as the minimum time expected directly from the time-energy uncertainty in Eq. \eqref{teu}.
Further, it appears from Eq. \eqref{tint} that 
the minimum time interval diverges as the energy approaches Planck scale ($E\sim 1/4\alpha$). However, this divergence could be unphysical since the Schr\"{o}dinger equation \eqref{schrod} is deformed to first order in $\alpha$ only. 
Finally, as expected, a continuous time is recovered in the limit in which $\alpha\to 0$. In short,
any physical system with finite energy can only evolve by taking discrete jumps in time rather than continuously.

\section{Rate of Spontaneous Emission}
We now apply the above to a concrete 
quantum mechanical system. 
The rate of spontaneous emission  in a two-level system is well understood \cite{griffith}. Here we shall repeat this analysis for a deformed quantum mechanical system. Now for
  a two-level system with eigenstates $\psi_a$ and $\psi_b$, the eigenvalues of the unperturbed Hamiltonian $H^0$ can be written as 
\begin{equation}
H^0\psi_a=E_a\psi_a, \qquad H^0\psi_b=E_b\psi_b.
\end{equation}
Any state can be written as a superposition of those eigenstates with the time dependence found in Eq.(\ref{psit})
\begin{equation}
\label{psit2}
\Psi(t)=c_a\psi_a e^{\frac{-it}{2\alpha\hbar}\left(1-\sqrt{1-4\alpha E_a}\right)}+ c_b\psi_b e^{\frac{-it}{2\alpha\hbar}\left(1-\sqrt{1-4\alpha E_b}\right)}.
\end{equation}
If a time-dependent perturbation $H'(t)$ was turned on, the wave function $\Psi(t)$ can still be expressed as the previous equation but with a time-dependent $c_a(t)$ and $c_b(t)$, and the goal is to solve for $c_a(t)$ and $c_b(t)$. This will also hold if the time evolution of the system is given by a deformed 
Schr\"{o}dinger equation. So, let us assume that this system actually evolves according to the deformed    time-dependent Schr\"{o}dinger equation, 
\begin{eqnarray}
H \psi &=&H^0\psi +H'(t)\psi  \nonumber \\ &=&   i\hbar\partial_t\psi+\hbar^2\alpha\partial_t^2\psi. 
\end{eqnarray}
Now    neglecting terms of order $\hbar\alpha$ and $\hbar^2\alpha$ for a two-level system, we obtain 
\begin{align}
& c_a H^0\psi_a e^{-i\epsilon_a t/\hbar} +c_bH^0\psi_b e^{-i\epsilon_b t/\hbar} +c_aH'\psi_a e^{-i\epsilon_at/\hbar} + c_bH'\psi_b e^{-i\epsilon_bt/\hbar} \nonumber\\
&= i\hbar\left(\dot{c}_a\psi_ae^{-i\epsilon_a t/\hbar} +\dot{c}_b\psi_be^{-i\epsilon_b t/\hbar}\right)+c_aE_a\psi_ae^{-i\epsilon_at/\hbar}+c_bE_b\psi_be^{-i\epsilon_bt/\hbar}.
\end{align}
To simplify that last expression, we defined
\begin{equation}
\epsilon_a=\frac{1}{2\alpha}\left(1-\sqrt{1-4\alpha E_a}\right), \qquad \epsilon_b=\frac{1}{2\alpha}\left(1-\sqrt{1-4\alpha E_b}\right). 
\end{equation}
It may be noted that in the limit $\alpha \to 0$, we obtain  $\epsilon_a \to E_a$ and $\epsilon_b \to E_b$.
The first two terms cancel the last two terms. Now taking the inner product with $\psi_a$ and solving for $\dot{c}_a$, we obtain 
\begin{equation}
\dot{c}_a=-\frac{i}{\hbar}\left(c_aH'_{aa}+c_bH'_{ab}e^{-i\omega_0 t}\right). 
\end{equation}
Here we have defined
\begin{eqnarray}
H'_{ij}&=&\langle\psi_i|H'|\psi_j\rangle,\nonumber \\ 
\label{omega0}
\omega_0&=&\frac{\epsilon_b-\epsilon_a}{\hbar}\nonumber \\ &=&\frac{\sqrt{1-4\alpha E_a}-\sqrt{1-4\alpha E_b}}{2\alpha\hbar}.
\end{eqnarray}
Similarly, the inner product with $\psi_b$ picks out $\dot{c}_b$, 
\begin{equation}
\dot{c}_b=-\frac{i}{\hbar}\left(c_bH'_{bb}+c_aH'_{ba}e^{i\omega_0 t}\right).
\end{equation}
Since in most applications the diagonal elements of $H'$ vanish, we get the simplified equations
\begin{equation}
\label{cacb}
\dot{c}_a=-\frac{i}{\hbar}H'_{ab}e^{-i\omega_0 t}c_b, \qquad \dot{c}_b=-\frac{i}{\hbar}H'_{ba}e^{i\omega_0 t}c_a.
\end{equation}

These equations have the same form as the un-deformed two-level system, except that  in these equations    $\omega_0$ is modified.  Thus, the standard analysis for the un-deformed two-level system also holds for a deformed two-level system.   So if   an atom is exposed to a sinusoidally oscillating electric field ${\bf E}= E_0\cos(\omega t)\hat{k}$, then the perturbation Hamiltonian can be written as 
\begin{equation}
H'(t)=-qE_0 {\bf r}\cos(\omega t)
\end{equation}
and
\begin{equation}
H'_{ba}=-{\bf p} E_0\cos(\omega t), 
\end{equation}
where  ${\bf p}=q\langle\psi_b|{\bf r}|\psi_a\rangle$ is the electric dipole radiation. Repeating the analysis for the un-deformed two-level system   \cite{griffith},  we can write the rate of spontaneous emission  $\mathcal{A}$  for the deformed system as 
\begin{equation}
\mathcal{A}=\frac{\omega_0^3|{\bf p}|^2}{3\pi\epsilon_0\hbar}.
\end{equation}
Expanding to first order in $\alpha$,  we obtain 
\begin{equation}
\label{emmrate}
\mathcal{A}=\frac{(E_b-E_a)^3 |{\bf p}|^2}{3\pi\epsilon_0\hbar^4}+\frac{(E_b-E_a)^3(E_a+E_b) |{\bf p}|^2}{\pi\epsilon_0\hbar^4}\alpha.
\end{equation}

To get an order of magnitude estimate of the effect of the extra term in Eq. \eqref{emmrate}, we consider the spontaneous emission from a transition between the first and second energy levels in the hydrogen atom. Now for these levels, we have $E_1=13.6$eV, $E_2=E_1/4$, and   $|{\bf p}|\sim 0.7qa_0$, where $a_0$ is the Bohr radius. Thus, we obtain 
\begin{align}
\mathcal{A} &\approx 2.1 + 1.7\times 10^{-17} \alpha \text{~[m}^{-1}]\\ \nonumber
&\approx 6.2 \times 10^8 + 5.1 \times 10^{-9} \alpha \text{~[s}^{-1}].
\end{align}
  The uncertainty in measuring the rate of  spontaneous emission  for hydrogen atom is $\pm 0.3 $ \% \cite{transp}.  So,  the bound on $\alpha_0$ 
from the rate of  spontaneous emission  in a  hydrogen atom is given by 
\begin{equation}
\alpha_0 < 7.2 \times 10^{23}.
\end{equation}
Hence, at this scale the effect of the rate of  spontaneous emission  in  hydrogen can be effected by the temporal  deformation proposed in this paper. If such a deformation scale exists at this scale in nature, future measurements might be able to detect it. 

It may be noted that we  can also use the lifetime of particles to set bounds on $\alpha_0$, for the modified Schr\"{o}dinger equation. For example, the tau has a lifetime of $(290.3\pm 0.5)\times 10^{-15}$ s  \cite{rpp}, and since the minimum time from Eq. \eqref{texp} must be less than the uncertainty in measuring the tau's life time, then
$ 2\pi\hbar\alpha < 0.5 \times 10^{-15}$. This means that $\alpha_0<1.5\times 10^{27}$. However,  the bound on $\alpha_0$ from the hydrogen atom is 
more stringent than the bound on $\alpha_0$ from the lifetime of particles. So, in the case that a minimum measurable time exists in nature,  we are more likely to   first observe  its  effects on the rate of  spontaneous emission  in hydrogen atoms.

\section{Conclusions}

We have shown here that
the existence of a minimum measurable time scale in a quantum theory naturally leads to the discretization of time. 
This is similar to the existence of a minimum measurable length scale leading to a  discretization of space.
Thus, a crystal in time gets naturally formed by the existence of a minimum measurable time scale in the universe. 
Time crystals have been studied recently for systems in which
time reparametrization is broken, just as
spatial translation is broken in regular crystals. 
Time crystals have also been studied earlier for analyzing superconducting rings \cite{ctime,  ctime1, c0time, c1time, cti}. We also analyzed the effect of such a deformation on the rate of  spontaneous emission  in a hydrogen atom.  
It would be interesting to analyze a combination of minimum length and minimum time 
deformations of quantum mechanics to demonstrate a discretization of space and time in four dimensions. 
We expect to obtain non-local fractional derivative 
terms in that case, which may possibly be dealt with
using a theory of harmonic extension of functions 
\cite{mf00, mf22}, or via the Dirac equation approach 
\cite{main2}. 
It may be noted that it is conceptually useful to view the minimum measurable time as
a component of a minimum Euclidean four volume with complex time, and then analytically continue 
the results  to  a Lorentz manifold. However, as we analyzed a  system with   Galilean symmetry,
we did not to go through this procedure. 

It is expected that the deformation of the Hamiltonian 
studied here will affect all physical systems. 
Thus for example, one can study the decay rates of particle and unstable nuclei using this deformed time evolution, which
are expected to change as well. In fact, by fixing the value of this deformation parameter just below the experimentally measured limit, it might be possible to devise tests for detecting such deformation of time evolution of quantum mechanics. 
The deformed Hamiltonian should affect 
time-dependent perturbation theory as well. 
For example, the out-of-equilibrium Anderson model has been studied using the time-dependent density functional theory  \cite{an}. This has important applications for time-dependent processes in an open system where different scattering processes take place. This behavior will get modified due to this deformation of quantum mechanics. Similarly the
quantum mechanical systems for which the strict adiabatic
approximation fails, but which do not escape too far from the adiabatic limit, can be analyzed using
a time-dependent adiabatic deformation of the theory \cite{ad}. 
It would be interesting to analyze the effect of having a minimum  measurable time for such a  time-dependent adiabatic deformation of the theory. 

\vspace{0.3cm}
\noindent
{\bf Acknowledgment}\\
The work of SD is supported by the Natural Sciences
and Engineering Research Council of Canada.

\end{document}